\documentclass[preprint,showpacs,preprintnumbers,amsmath,amssymb,nofootinbib]{revtex4}
\input psfig.sty
\def \be {\begin{equation}}
\def \ee {\end{equation}}
\def \bea {\begin{eqnarray}}
\def \eea {\end{eqnarray}}

\usepackage{graphicx}
\usepackage{dcolumn}
\usepackage{bm}

\begin{document}

\title{Power-law statistics and stellar
rotational velocities in the Pleiades}

\author{J. C. Carvalho$^{1}$}
\email{carvalho@dfte.ufrn.br}

\author{R. Silva$^{1,2}$}
\email{raimundosilva@dfte.ufrn.br}

\author{J. D. Jr. do Nascimento$^{1}$}
\email{dias@dfte.ufrn.br}

\author{J. R. De Medeiros$^{1}$}
\email{renan@dfte.ufrn.br}

\affiliation{$^{1}$Universidade Federal do Rio Grande do Norte, UFRN, Departamento de F\'{\i}sica
C. P. 1641, Natal - RN, 59072-970, Brazil}

\affiliation{$^{2}$Universidade do Estado do Rio Grande do Norte, 59610-210, Mossor\'o, RN, Brasil}


\date{\today}

\begin{abstract}
In this paper we will show that, the non-gaussian statistics
framework based on the Kaniadakis statistics is more appropriate
to fit the observed distributions of projected rotational velocity
measurements of stars in the Pleiades open cluster. To this end,
we compare the results from the $\kappa$ and
$q$-distributions with the Maxwellian.
\end{abstract}

\maketitle

\section{Introduction}


Some restrictions to the applicability of the statistical mechanics have motivated the investigation of the power-law or non-gaussian statistics, both from theoretical and experimental viewpoints. In this concern, the nonextensive statistical mechanics \cite{tsallis1} and extensive generalized power-law statistics \cite{k1} are the most investigated frameworks.
In the later one, recent efforts on the kinetic foundations of the $\kappa$-statistics leads to a power-law
distribution function and the $\kappa$-entropy which emerges in the context of the special
relativity and in the so-called {\it kinetic interaction principle}
(see Ref. \cite{k1}). Formally, the $\kappa$-framework is based on
$\kappa$-exponential and $\kappa$-logarithm functions, which are
defined by
\begin{equation}\label{expk}
\exp_{\kappa}(f)= (\sqrt{1+{\kappa}^2f^2} + {\kappa}f)^{1/{\kappa}},
\end{equation}
\begin{equation}\label{expk1}
\ln_{\kappa}(f)= {{f^{\kappa}-f^{-\kappa}\over 2\kappa}},
\end{equation}
whereas the $\kappa$-entropy associated with the $\kappa$-statistics is given by
\cite{k1}
\begin{equation}\label{first}
S_\kappa = - \int d^3 p f\ln_\kappa f =  - \langle{\ln_\kappa
(f)\rangle}.
\end{equation}
The expressios above reduces to the
standard results in the limit $\kappa=0$.

The Tsallis statistics has been investigated in a wide range of
problems in physics \footnote{For a complete and updated list of
refences see http://tsallis.cat.cbpf.br/biblio.htm}. In the
astrophysical domain, the first applications of this powerlaw
statistics studied stellar polytropes \cite{plastino93} and the
peculiar velocity function of galaxy clusters \cite{lavagno98}.
More recently, Kaniadakis statistics has also been studied in the
theoretical and experimental context \cite{everybody2}, however
the first application with a possiple connection with
astrophysical system has been the simulation in relativistc
plasmas. In this regard, the powerlaw energy distribuition
provides a strong argument in favour of the Kaniadakis statistics
\cite{lapenta07}. Therefore, in such systems where nonextensivity
holds, the classical statistics may be generalized. In this sense,
one of the most puzzling questions in stellar astrophysics in the
past 50 years is that concerning the nature of the statistical law
controlling the distribution of stellar rotational velocity, in
spite of the large acceptance that stellar rotation axes have a
random orientation \cite{struve45}. In the middle of the past
century, Chandrasekhar and Munch \cite{munch50} were the first to
derive analytically the distribution of stellar projected
rotational velocity, on the basis of a Gaussian distribution. In
their approach, these authors first assumed a parametric form for
a function $f(v)$, where $v$ is the true rotational velocity, then
computed the corresponding distribution of the projected
rotational velocity $V sin i$ and finally adjusted a set of
stellar parameters to reproduce the $V sin i$ measurements. Two
decades later, Deutsch \cite{deut70} claimed that the distribution
of stellar rotational velocities should have the form of a
Maxwellian-Boltzmann law. Nevertheless, a number of studies have
shown a clear discrepancy between theory and observations, where
observed distributions are not fitted by a Gaussian or Maxwellian
function with a good level of significance. A Gaussian or
Maxwellian distribution that fits the fast rotators fails to
account for low rotation rates. On the other hand, a fit to slow
rotators fails to explain the rapidly rotating stars
\cite{dwo74,Wolff82}.

Rotation is one of the most  important observable in stellar
astrophysics, driving strongly the evolution of stars, providing
also valuable informations on stellar
 magnetism, mixing of chemical abundances in stellar interior, tidal
interaction in close binaries, and engulfing of brown dwarfs and planets.
In addition, if the present value of the rotational velocity of stars at a given
evolutionary stage reflects the original angular momentum with which they
were formed, the behaviour of the distribution of rotational velocity may also
be used to study some of the characteristics of the physical processes
controlling star formation.  Early studies on the nature of the statistics controlling the distribution of stellar
rotational velocity  were based on $V sin i$ measurements with poor precision, which, admittedly,
can lead to systematic errors on the final analyses for
low $V sin i$ values.  Here, we show that the question of the nature
of the distribution of stellar rotational velocity,
at least for low-mass stars in the Pleiades open cluster, is not
simply a question of which mathematical
function model is used, but it depends primarily on the
statistical mechanics applied, which should be general
enough to take into account the changes in rotation with time.

In this work we have investigated the effects of the powerlaw
statistics on the observed distribution of projected rotational
velocity measurements of stars in the Pleiades open cluster by
considering a $\kappa$-distribution functions. In this regard, it
is worth emphasis that earlier study based on the Tsallis
statistics has been developed in Ref. \cite{soares06}. However, in
the present work, to study the effects of the powerlaw statistics
on the observed distribution in the Pleiades open cluster we use
the more recent generalization of Kaniadakis \cite{k1} and, for
completeness, we compare the results with the ones obtained in the
context of the Tsallis statistics. This paper is organized as
follows. In Sec II, based on the basic formalism presented in Ref.
\cite{deut70}, we present a generalization of the rotational
velocity distribution in the spirit of Kaniadakis statistics. A
brief discussion on the stellar sample is made in Sec. III. Our
main results are discussed in Sec. IV and we summarize the main
conclusions in Sec. V.

\section{$\kappa$-distribution function}

As well know, a large portion of the experimental evidence, as
well as some theoretical considerations supporting Tsallis and
Kaniadakis proposal involves a non-Maxwellian (powerlaw)
distribution function associated with the thermostatistical
description from the variety of the physical systems
\cite{everybody1,everybody2}. In Tsallis framework, the
equilibrium velocity $q$-distribution may be derived from at least
three different methods, namely: (i) through a simple nonextensive
generalization of the Maxwell ansatz, which is based on the
isotropy of the velocity space \cite{silva98}; (ii) within the
nonextensive canonical ensemble, that is, maximizing Tsallis
entropy under the constraints imposed by normalization and the
energy mean value \cite{abe99} and (iii) using a more rigorous
treatment based on the nonextensive formulation of the Boltzmann
$H$-theorem \cite{lima01}. Here, we revisit the first method by
considering the Kaniadakis statistical which is based on
$\exp_\kappa$ and $\ln_\kappa$ given by Eq. (\ref{expk}), where
$f$ is a function of random variables that includes the standard
exponential as the limiting case when $\kappa\rightarrow 0$.

It is widely known that Deutsch \cite{deut70} has considered the
distribution function for the magnitude of a vector that has
random orientation. For this, it is required to find the
distribution function of a positive scalar $\omega$, which is the
magnitude of a vector ${\vec\omega}$. We assume that the
distribution of ${\vec\omega}$ is isotropic. We also assume that if
it is decomposed into components along Cartesian axes, the
distribution of any component is independent of the other
components. Deutsch has defined $\Omega$ as the non-dimensional
quantity $\omega j$, where $j$ is a parameter with the dimension
$\omega^{-1}$, so
\begin{equation}
{\vec{\Omega}}=\Omega_x {\vec i} + \Omega_y {\vec j} + \Omega_z {\vec k}
\end{equation}
The probability that $\Omega_x$ lies in the interval
$[\Omega_x;\Omega_x + d\Omega_x]$, $\Omega_y$ in
$[\Omega_y;\Omega_y + d\Omega_y]$ and $\Omega_z$ in
$[\Omega_z;\Omega_z + d\Omega_z]$ is then
\begin{equation}\label{eqgood}
F(\Omega)d^3\Omega=f(\Omega_x)f(\Omega_y)f(\Omega_z)d\Omega_xd\Omega_yd\Omega_z
\end{equation}
where $\Omega=\sqrt{\Omega_x^2+\Omega_y^2+\Omega_z^2}$ and $F(\Omega)$ is the standard Maxwellian distribution function
\begin{equation}
 F(\Omega)={4\over\sqrt{\pi}}\Omega^2\exp{(-\Omega^2)}
\end{equation}

Let us now consider the arguments given in Refs. \cite{silva98,
soares06}. One can modify the basic hypothesis of statistical
independence between the distributions associated with the
components of $\vec{\Omega}$, based on the $\kappa$-statistics. As
pointed out in \cite{silva98} and \cite{soares06}, the
independence between the three velocity components does not hold
in systems with long-range interaction, or statiscally correlated,
where the power-law statistics character is observed. Taking  such
arguments into account, the generalization for Eq. (6) in the
light from Kaniadakis the framework reads,
\begin{equation}\label{good1}
F(\Omega)d^3\Omega=\exp_{\kappa}(\ln_\kappa f(\Omega_x)+\ln_\kappa f(\Omega_y)+ \ln_\kappa f(\Omega_z))d\Omega_xd\Omega_yd\Omega_z,
\end{equation}
where the $\kappa$-exponential and $\kappa$-logarithm are given by identities (1) and (2). In particular, in the limit $\kappa=0$ the standard expression (\ref{eqgood}) is recovered. Note also that $\ln_\kappa(\exp_\kappa(f))=\exp_\kappa(\ln_\kappa(f))=f$, and ${d\ln_\kappa f\over dx}={f^{\kappa}+f^{-\kappa}\over 2f}{df\over dx}$ are satisfied. Therefore, the partial differentiation of the $\kappa$-ln of (\ref{good1}) with respect to $\Omega_i$ leads to
\begin{equation}
{\partial \ln_\kappa F\over\partial\Omega_i}={\partial\over\partial\Omega_i}(\ln_\kappa f(\Omega_x)+\ln_\kappa f(\Omega_y)+ \ln_\kappa f(\Omega_z)),
\end{equation}
or, equivalently,
\begin{equation}\label{above}
{F^{\kappa}+F^{-\kappa}\over 2F} {1\over\chi}{dF\over d\chi}={1\over\Omega_i}{d\over d\Omega_i} \ln_\kappa f_i,
\end{equation}

where $\chi=\sqrt{\Omega_x^2+\Omega_y^2+\Omega_z^2}$. Now, by defining
\begin{equation}
\phi(\chi)={F^{\kappa}+F^{-\kappa}\over 2F} {1\over\chi}{dF\over d\chi},
\end{equation}
we may rewrite (\ref{above}) as
\begin{equation}\label{above1}
\phi(\chi)= {1\over\Omega_i}{d\over d\Omega_i} \ln_\kappa f_i
\end{equation}
The second member of the above equation depends only on
$\Omega_i$, with  $i=x,y,z$. Hence, equation (\ref{above1}) can be
satisfied only if all its members are equal to one and the same
constant, not depending on any of the velocity components. Thus,
we can make $\phi(\chi)=-2/\sigma_{\kappa}^2$, where the parameter
$\sigma_{\kappa}$ is the width of the $\kappa$-distribution,
leading to
\begin{equation}\label{newanzatz}
 {1\over\Omega_i}{\partial\over\partial\Omega_i}(\ln_\kappa f_i)=-{2\over\sigma_{\kappa}^2}.
\end{equation}
Hence, the solutions of Eq. (\ref{newanzatz}) for $f(\Omega_i)$ is
given by the $\kappa$-distribution
\begin{equation}\label{kappadist}
 f(\Omega_i)= \exp_{\kappa} (-\Omega^2_i / \sigma_{\kappa}^2)
\end{equation}
From (\ref{kappadist}) we see that the Gaussian probability curve
is replaced by the charactheristic powerlaw behavior of Kaniadakis
framework and, as expected, the limit $\kappa = 0$ recovers the
exponential result. Note also that for any values of the
$\kappa$-parameter, the powerlaw (\ref{kappadist}) does not
exhibits a cut-off in the maximal allowed rotational velocities,
and it is straightforward to show that $F(\Omega)$ is given by
\begin{equation}
 F(\Omega)= \exp_{\kappa} (-\Omega^2 / \sigma_{\kappa}^2) 
\end{equation}

The probability of finding $\omega$ in the interval $[\Omega,\Omega+d\Omega]$ is determined
by $\Psi(\Omega)=\int f(\Omega)d^3\Omega$ which leads to a power law that belongs to the
  same class of power law as
given in Eq. (\ref{kappadist}), i.e.,
\begin{equation}
 \Psi(\Omega)= 4\pi \Omega^2 \exp_{\kappa} (-\Omega^2 / \sigma_{\kappa}^2).
\end{equation}
 Here, it is worth mentioning that the standard distribution of the true rotational
 velocity $V$ for a star sample is $F(V)\sim V^2\exp(-V^2)$. As shown
by Deutsch \cite{deut70}, the standard observed distribution of the projected rotational
velocity $V sin i$, for a random orientation of axes, must be given by
$\phi(y)\sim y\exp(-y^2)$ \cite{Kraft70}, with $y = V sin i$. Henceforth, the
$\kappa$-distribution $\phi_\kappa(y)$ should reproduce the standard one,
in the same way as $F_\kappa(v)$ recovers $F(V)$ in the $\kappa=0$ limiting case.
Therefore, by considering this arguments, we introduce the following distribution
function for the observed stellar rotational velocities
\begin{equation}
\label{kappadistY}
 \phi_\kappa(y)= y  \exp_{\kappa} (-y^2 / \sigma_{\kappa}^2).
\end{equation}

\begin{table}[tbp]
\begin{center}
\caption[]{Best values of the parameters of Kaniadakis ($\kappa$ and $\sigma_{\kappa}$)
and Tsallis ($q$ and $\sigma_q$)
distribution determined using Kolmogorov--Smirnov test for the
rotational velocity of stars in the Pleiades cluster.}
\label{tab0} \vspace{0.4in}
\begin{tabular}{cccccccc}
\hline\hline
$\Delta(B-V)$ &    N   &  $\kappa$   &  $\sigma_{\kappa}$  & $P_{max}$ & $q$   &  $\sigma_q$  & $P_{max}$  \\
 \hline
 \hline
&&&&&&&\\
0.40--1.40 & 217 & $0.446^{+0.048}_{-0.073}$ & 7.81 & 0.28 & $ 1.334^{+0.038}_{-0.055}$ &  6.93 & 0.23 \\
&&&&&&&\\
\hline
\hline
\end{tabular}
\end{center}
\end{table}
\begin{figure}[t]
\vspace{.2in}\centerline{\psfig{figure=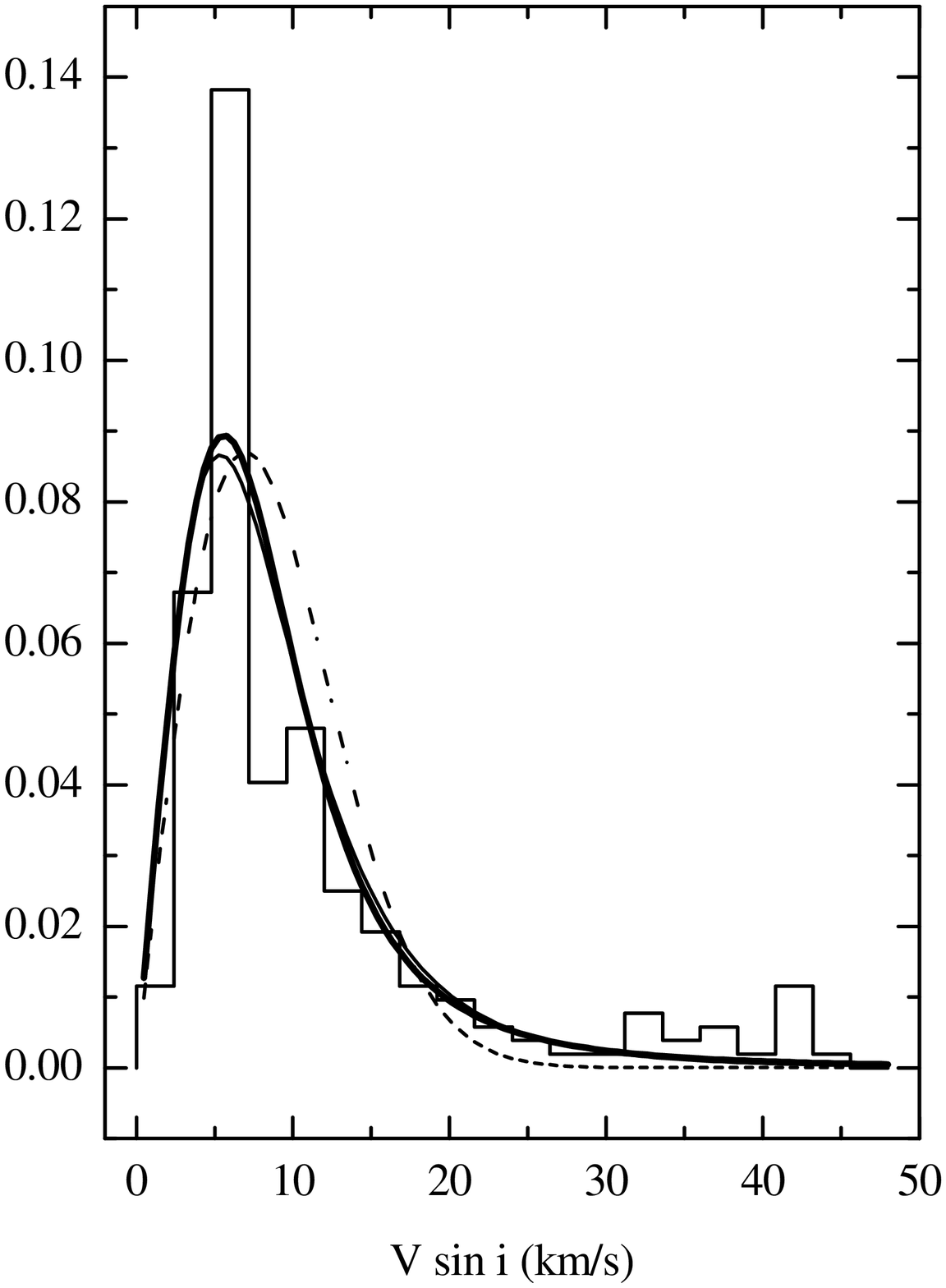,width=4.0truein,height=4.5truein}\hskip 0.1in}
\caption{Observed distribution (histogram) of the rotational
velocity of the stars in the Pleiades open cluster.
The curves represent the best fitted
maxwellian (dashed line), Tsallis (thin line) and Kaniadakis
(thick line) distribution. The fitting parameters are in Table 1.}
\label{fitA}
\end{figure}

\section{The stellar sample}
\label{sample}

The rotational velocities $V\sin{i}$ used in the present analysis
were taken from the rotational survey for the Pleiades stars
carried out by Queloz et al. \cite{Queloz98}. We have selected 219 stars from
the original sample given by those authors. All the selected
objects are low-mass stars and provide a complete and unbiased
rotation data set for stars in the $B-V$ range ($0.4-1.4$)
corresponding to an effective temperature range from
4000 K to 6000 K  and a mass range from $0.6 M_{\bigodot}$ to $1.2
M_{\bigodot}$. For a complete discussion on the observational
procedure, calibration and error analysis the reader is referred
to Queloz et al. \cite{Queloz98}. We should observe that
individual errors in $V\sin{i}$ measurements are better than about
1 km/s and should not play a significant role on the observed
distributions.

If we plot the observed distribution we note that there are two stars
with velocity (105 km/s and 160 km/s) too far from the peak
of the distribution (5 -- 8 km/s). We have, therefore, excluded these stars with
exceedingly high value of $V\sin{i}$ from the sample.

In order to avoid biases due to arbitrary choices of bin range when
constructing the frequency histograms, we have
decided to study the observed cumulative distribution of the
rotational velocities, $V\sin{i}$, and compare it
with the integral of the probability distribution function in
(\ref{kappadistY}). The normalized cumulative distribution is
given by
\begin{equation}
\Sigma_\kappa(y)= {\int_0^y  y  \exp_{\kappa} (-y^2 / \sigma_{\kappa}^2) dy  \over \int_0^\infty  y  \exp_{\kappa} (-y^2 / \sigma_{\kappa}^2) dy }.
\end{equation}

\section{Results}

\begin{figure}
\vspace{.2in}\centerline{\psfig{figure=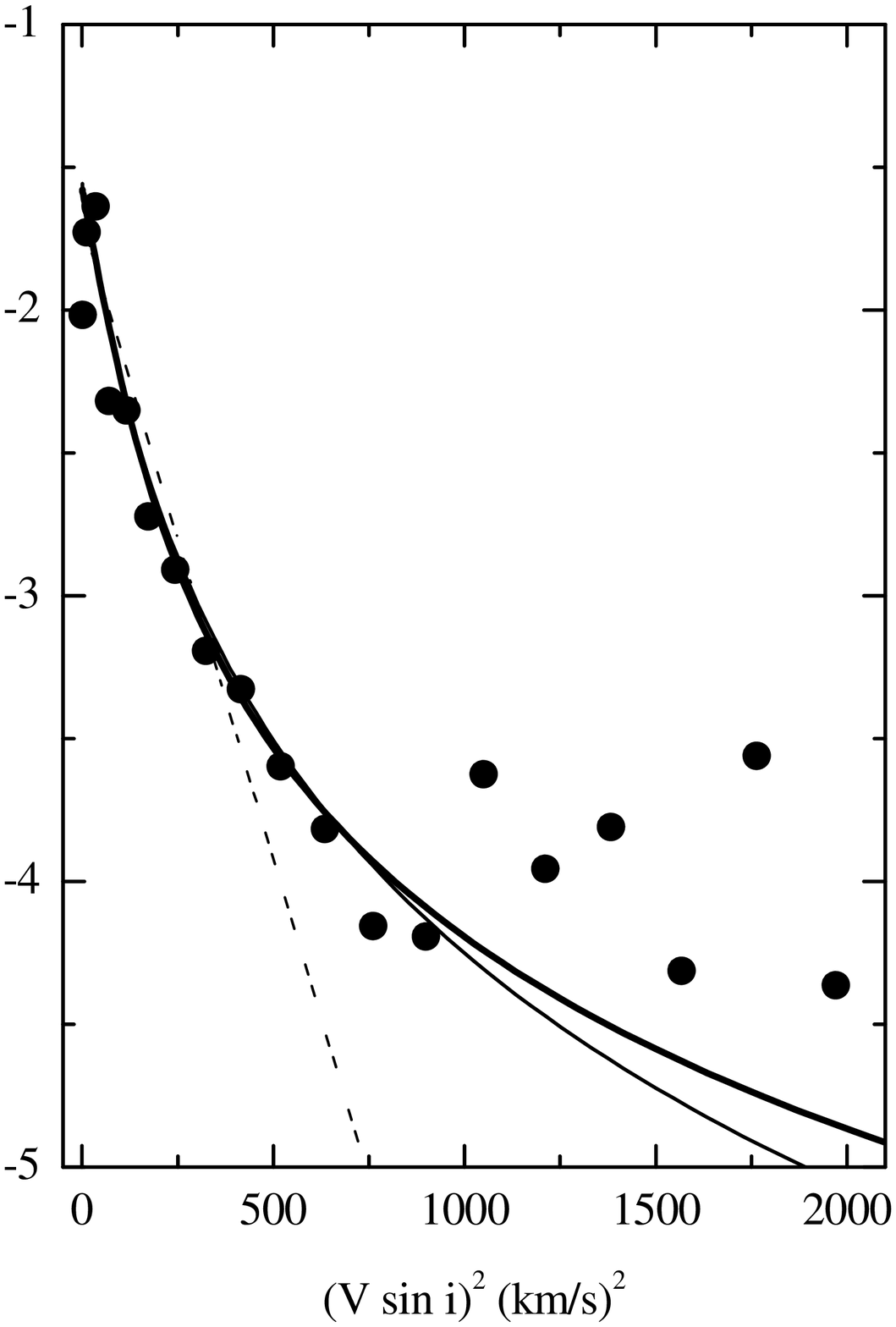,width=4.0truein,height=4.5truein}\hskip 0.1in}
\caption{As in Fig. \ref{fitA} but for the logarithm of the
distribution function divided by $V\sin{i}$ as a function of
$(V\sin{i})^2$.}
\label{fitlogA}
\end{figure}

Parallel with the Kaniadakis distribution we have also fitted the
data to the Tsallis distribution in order to evaluate which one
would best fit the observations. The Tsallis generalized
maxwellian is a two parameter ($q$ and $\sigma_q$) function given
by
\begin{equation}
\label{qdistY}
 \phi_q(y)= y \left[1-(1-q){y^2\over\sigma_q^2}\right]^{1/(1-q)}.
\end{equation}
In the limit $q=1$ the standard Maxwellian is recovered.

To calculate the best values of the distributions parameters
for the observed cumulative distribution
we used the Kolmogorov-Smirnov statistical test.

The distribution functions were used to fit the observational data,
to obtain the best $\phi_\kappa(y)$ and $\phi_q(y)$, giving
the best $\kappa$ and $q$-value together with the best $\sigma_{\kappa}$ and
$\sigma_q$ for the corresponding distribution. The results are
shown in Table 1. We can clearly see that the $V\sin{i}$ distribution of the Pleiades stars
do not obey a standard Maxwellian function since the values of
$\kappa$ and $q$ are significantly different from 0 and 1,
respectively.

Fig. \ref{fitA} shows the best fits for the histogram of the
observed distribution of $V\sin{i}$ according the results in Table 1.
The $\kappa$ and q-Maxwellian functions are represented, respectively
by the thick ($\kappa=0.45$) and thin ($q=1.33$) lines. The dashed line
represents standard Maxwellian function.  The
distribution of observed $V\sin{i}$ is without a doubt more adequately
fitted by either the $\kappa$ or the q-Maxwellian function. This is more
noticeable in Fig. \ref{fitlogA} where we have plotted the logarithm of the
distribution divided by $V\sin{i}$, that is $log(\phi(y)/y)$, as a
function of $(V\sin{i})^2$ so that the standard Maxwellian is
represented by a straight (dashed) line. The Kaniadakis
distribution is represented by the thick line while Tsallis
distribution by the thin line. We observe that none of the distribution
fit well the high velocity end of the observed data. Although the Kaniadakis
function fits the data slightly better than Tsallis
function, the difference may be regarded as marginal as indicated by
the values of the maximum probability of the Kolmogorov-Smirnov
statistical test (Table 1).

\begin{figure}[t]
\vspace{.2in}\centerline{\psfig{figure=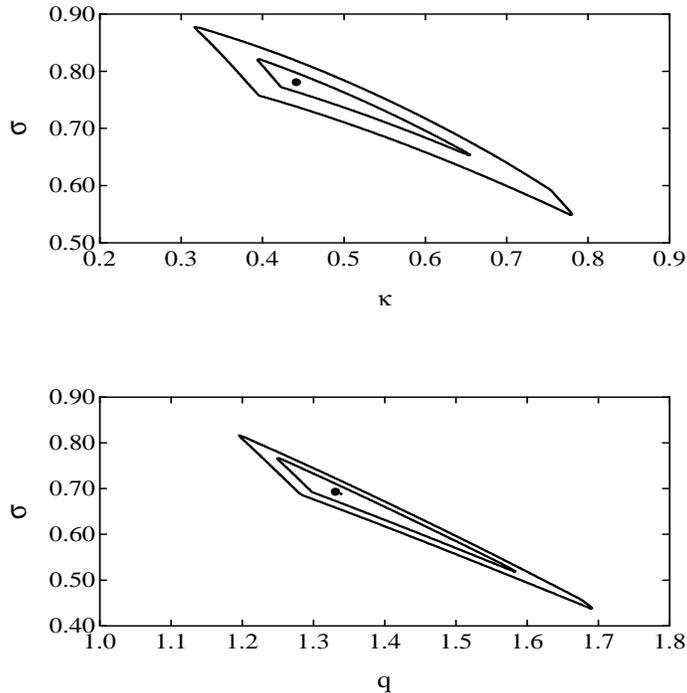,width=4.0truein,height=4.5truein}\hskip 0.1in}
\caption{Rejection region (outside curves) of the null hypothesis
that the $V \sin{i}$ distribution is drawn from the
$\kappa$-distribution (upper panel) at 0.05 and 0.15 confidence level and
from $q$-distribution (lower panel) at 0.05  and 0.10 confidence level.
The dot ($\bullet$) represents the maximum probability for the pair
($\kappa$ -- $\sigma_{\kappa}$) and ($q$ -- $\sigma_{q}$).} \label{KT}
\end{figure}

Finally, in Fig. \ref{KT} we present the behaviour the parameter $\sigma$, representing
the width of the two non-Maxwellian, as a function of the parameter $\kappa$ and $q$.
It is clear that, at least for the present stellar sample, the
standard Maxwellian ($\kappa=0$ or $q=1$) is in the rejection region, i.e.,
outside the curve which delineates 0.05 significance level.

\section{Conclusions}

In this work we have used non-gaussian statistics to investigate the
observed distribution of projected rotational velocity
of stars in the Pleiades open cluster. We have studied in details the
Kaniadakis distribution and shown that it fits more closely the observed distribution
than the standard maxwellian. A comparison with the Tsallis
statistics shows that, at least when the observed distribution presents an
extended tail, as it is the case of rotational velocity
of stars in the Pleiades, both distributions give equivalent,
though not entirely satisfactory results.

As discussed in Sec. IV, the best fits for  the histogram of
observed distribution are non-gaussian with $\kappa=0.45$ and
$q=1.33$ for Kaniadakis and Tsallis parameter, respectivily. In
particular, we emphasize that the result of the $q$-parameter is
consistent with the upper limit $q<2$ obtained from several
independents investigations in the quantum limit \cite{abe03} and
in the non-quantum limit \cite{final}.

Finally, it is worth mentioning that the the best fits in $\kappa$
and $q$ can be recalculated by considering a more robust  stellar
sample. In this respect, the stellar radial velocity of a sample
of open clusters are being studied. This issue will be addressed
in a forthcoming communication.

\vspace{0.5cm}

\noindent {\bf Acknowledgments:} The authors are partially supported by the Conselho Nacional
de Desenvolvimento Cient\'{\i}fico e Tecnol\'ogico (CNPq -
Brazil). J. R. de Medeiros acknowledges financial support of the FAPERN Agency.

\end{document}